%% file: sigkddExp.tex
\documentclass{sigkddExp}
\usepackage{lipsum}
\usepackage{mathtools}
\usepackage[multiple]{footmisc}
\usepackage{graphicx}
\usepackage{subcaption}
\usepackage[export]{adjustbox}
\usepackage{hyperref}
\usepackage{caption}
\usepackage{url}
\usepackage{balance}
\hypersetup{
	colorlinks,
	linkcolor={blue},
	citecolor={blue},
	urlcolor={blue}
}
\begin{document}\sloppy
	\title{Intelligent Disaster Response \\ via Social Media Analysis - A Survey}
	\author{
		\alignauthor Tahora H. Nazer\textsuperscript{*}, Guoliang Xue\textsuperscript{*}, Yusheng Ji\textsuperscript{\dag}, and Huan Liu\textsuperscript{*}\\
		\affaddr{\textsuperscript{*}Arizona State University, USA\\
			\textsuperscript{\dag}National Institute of Informatics, Japan}\\
		\email{\{tahora.nazer,xue,huan.liu\}@asu.edu and kei@nii.ac.jp}
		\alignauthor Huan Liu\\
		\affaddr{Arizona State University}\\
		\email{huan.liu@asu.edu}
	}
	%\date{\date{}}
	\maketitle
	%\tableofcontents
	\begin{abstract}
\input{abstract.tex}

	\end{abstract}
%	\clearpage
	\section{Introduction}
	\input{intro.tex}

	\section{Characteristics of Disasters}
	\input{characteristics.tex}

	\section{Data Extraction and Filtering}
	\input{data.tex}
	\subsection{Data Extraction}
	\input{data_extraction.tex}
	\subsection{Data Filtering}
	\input{data_filtering.tex}	
		
	\section{Warning via Social Media}
	\input{warning.tex}
	
	\section{Reflection of Disaster Impact on Social Media}
%	\subsection{Language Change}6
	\input{impact.tex}

	\section{Facilitating Response via Social Media}
	\input{response.tex}

	\subsection{Tracking Disasters}
	\input{tracking.tex}
	
	\subsection{Situational Awareness}
	\input{situational_awareness.tex}

	\section{Relief Assistance via Social Media}
	\input{relief.tex}
	
	\section{Conclusion}		
	\input{conclusion.tex}

	\section{Looking Ahead}
	\input{future.tex}
	
	\section{Acknowledgments}
	This research was supported, in part, by NSF grant 1461886, ONR grant N00014-16-1-2257,and JST Strategic International Collaborative Research Program (SICORP). The information reported here does not reflect the position or the policy of the funding agencies. The authors would like to thank members of Data Mining and Machine Learning (DMML) Lab at Arizona State University and Lei Zhong from National Institute of Informatics, Japan for their feedback and contributions.
	\balance
	\bibliographystyle{abbrv}
	\bibliography{sigkddreferences}
\end{document}

%% file: abstract.tex
The success of a disaster relief and response process is largely dependent on timely and accurate information regarding the status of the disaster, the surrounding environment, and the affected people. This information is primarily provided by first responders on-site and can be enhanced by the first-hand reports posted in real-time on social media. Many tools and methods have been developed to automate disaster relief by extracting, analyzing, and visualizing actionable information from social media. However, these methods are not well integrated in the relief and response processes and the relation between the two requires exposition for further advancement. In this survey, we review the new frontier of intelligent disaster relief and response using social media, show stages of disasters which are reflected on social media, establish a connection between proposed methods based on social media and relief efforts by first responders, and outline pressing challenges and future research directions.

%% file: intro.tex
% DISASTERS
%Disasters are either natural such as hurricanes, floods, and earthquakes or man-made like bombing and political unrests. In response to disasters, relief agencies, such as Red Cross and Red Crescent, provide humanitarian assistance to the affected people; such efforts include food, shelter, health services, emotional support, relocation, and repairing/rebuilding the failed constructions\footnote{\texttt{http://goo.gl/jRM41u}}. Along with the on-site efforts, social media is being used as a tool for warning, informing, and responding to disasters.

%Based on a survey performed by Red cross in 2011, 80\% of people are willing to share photos and videos of disasters on social media and 1/3 of social media users would ask for help on social media~\cite{national2013public}.

%Social media has been the first venue in which the information of London subway bombing~\cite{crowe2012disasters} and Virginia Tech shootings\footnote{\texttt{http://goo.gl/1NUSuP}} were reflected and traditional news media exploited them in their reports.

% SM vs traditional sources
Social media is a new way of communication in the course of disasters. A major difference between social media and traditional sources is the possibility of receiving feedback from the affected people. Responders such as Red Cross can benefit this two-way communication channel to inform people and also gain insight by monitoring their posts. Twenty million tweets after Hurricane Sandy (2012) and eight million tweets after the Boston Marathon Bombings (2013)~\cite{gupta20131} have been published on Twitter. This swarm of posts can provide valuable insight and help with the disaster management when the functioning of a community is disrupted due to severe fatalities and infrastructural damage~\cite{disasterandemergency,smith2013environmental}. 

%Insights provided by SM in disasters
Two types of insight can be obtained from social media in the course of disasters. The ``big picture'' is an estimate of the scope of the disaster: area, casualties, and failed structures. ``Insightful information'' is more detailed and is available when more data is available on social media. Locations that need food, medical supplies, or blankets are examples of insightful information~\cite{palen2016crisis}.

%Challenges in getting insight from social media: social science and computer science
%Challenges in using social media for disaster response are two-fold. The first one is the separation between the issues researched in social sciences and the solutions proposed by computer science. In the former, disasters and their effect in on people is studied. Also temporal and social aspects of disasters are analyzed and eight stages for disasters are considered: pre-disaster, warning, threat, impact, inventory, rescue, remedy, and recovery~\cite{powell1954introduction}. Each stage has specific characteristics and requires different actions by disaster responders. Computer science research, on the other hand, monitors a disaster through the lens of people's activity in the online world and specifically social media. Numerous challenges have been addressed by research on social media, however, the relation between these solutions and their applicability and relation to the status of disasters has not been thoroughly considered.

% First challenge in using SM in disasters
One challenge associated with acquiring insight via social media is processing enormous amount of information in a timely manner. After Japan earthquake and tsunami (2011), 1,200 tweets were published every minute from Tokyo~\cite{url2011Twitter} and after Hurricane Sandy (2012), the peak rate of 16,000 tweets per minute has been reported~\cite{meier2015digital}. This amount of data is too large to be manually processed by emergency responders. Some of the proposed methods to overcome this issue are presented in Section~\ref{sec:data_extraction}.
%Moreover this data is mixed with unwanted content such as spam and rumors.

%To automate this process several methods have been proposed to first assure that as many related posts to the target event as possible is collected (see Section~\ref{sec:extraction}) and then specific sets of posts such as informative tweets are extracted (see Section~\ref{sec:situational_awareness}). 

% Second challenge in using SM in disasters
Another challenge is the wide-spread of unwanted content such as daily chatter, spam, and rumor in social media. Among the 8 million tweets related to Boston Marathon Bombings (2013), 29\% were found to be rumors and 51\% to be generic opinions and comments~\cite{gupta20131}. Moreover, exploiting bots has worsened this issue. Large number of bots can be generated in a short period of time and be used to spread spam, deviate the conversation of real users, and help with the virality of rumors. We mention some of the solutions to this challenge in Section~\ref{sec:data_filtering}.

Disasters have eight socio-temporal stages: Pre-disaster, Warning, Threat, Impact, Inventory, Rescue, Remedy, and Recovery~\cite{powell1954introduction}.  The volume of social media posts varies in each stage; majority of users start posting after the disaster onsets and the frequency decreases when the disaster reaches its final stages. Availability of data is a major factor in building automatic methods for facilitating the management tasks. Hence, we consider four stages in disasters; the ones in which social media posts are dense enough for Machine Learning methods to achieve reliable results: Warning, Impact, Response, and Recovery.
%From another perspective, disaster management lifecycle stages are Hazard Analysis, Vulnerability Analysis, Mitigation \& Prevention, Preparedness, Prediction \& Warning, Response, and Recovery~\cite{aim1986schramm}.

%The details of each stage is explained in Section~\ref{sec:specs}

In the warning stage (Section~\ref{sec:warning}), social media can be used as a complementary source of information to help increase the confidence in predicting disasters and providing warnings. Changes in the frequency of posts with specific words and topics, activity patterns of users~\cite{asur2010predicting,sampson2015real,tumasjan2010predicting}, and sentiment of posts~\cite{mishne2006predicting} are used to predict disasters. Predicting disasters before they hit an area provides the opportunity to warn people in danger and evacuate elevators and operation rooms. Currently, USGS uses tweets to check the accuracy of sensor reports and detect earthquakes in a shorter time. Earthquakes can be detected using tweets by 60 seconds earlier than sensors; this time is valuable for warning areas in danger and starting evacuation processes~\cite{Ellis2015Twitter}.

When disasters impact an area, social media posts show anomalies such as changes in the language. A study~\cite{cohn2004linguistic} on LiveJournal after 11 September 2001 shows that emotional positivity decreased and cognitive processing, social orientation, and psychological distancing increased after the attack~\cite{beigi2016overview}. These changes in social media posts can be quantitatively captured in sentence level or topic level. Capturing the change in real-time results in detecting disasters before they are announced by official sources, governmental websites, or major news outlets~\cite{Sambuli2013useful} (see Section~\ref{sec:impact}).

In response to the chaotic environment caused by disasters, emergency responders want to acquire actionable insight and a big picture of the disaster~\cite{castillo2016big}. Detecting and tracking topics, trends, and memes on social media provides information regarding the status of disasters and the affected people. Damages, casualties, missing animals, and failed structures are some of the topics that people discuss on social media. Tracking these topics, discovering the trends, and monitoring mentioned locations help responder distribute resources more efficiently (see Section~\ref{sec:response}).

Volunteers are significantly important in the relief process. They post information that increases situational awareness (e.g. status of roads and damages to built structures) and provide technical support for translating social media posts and geotagging them. Some of the systems that exploit social media posts to facilitate disaster management are Ushahidi~\cite{okolloh2009ushahidi}, AIDR~\cite{imran2014aidr}, and TweetTracker~\cite{kumar2011tweettracker} which will be discussed in Section~\ref{sec:relief} with more details.

In this paper, we clarify the relation between the stages of disasters and relevant research on social media. This effort is towards unwrapping the potentials of social media to be exploited by emergency responders to a larger extent. We consider four stages for disasters which are widely reflected on social media: warning, impact, response, and relief. In each stage, we introduce approaches that use social media to ease the relief efforts. 
The major difference between this work and previous surveys in the field is the organization of the material in an effort toward facilitating the exploitation of these methods by disaster responders. We use disaster management stages used by first responders~\cite{aim1986schramm} to explain limitations and potentials of social media research. We bold available methods and tools that can be used by responders and mention the areas in which social media has not been used to its potential. Moreover, we focus on more recent areas which are not widely reflected in previous studies. We believe that this work establishes a connection between available tools based on social media data and the efforts of first responders.

Contributions of this paper are as follows:
\begin{itemize}
	\item Introducing four stages for disasters based on activities of users on social media.
	\item Categorizing research on social media for disaster management based on their application in each stage.
	\item Connecting the research on social media with disaster management efforts by first responders.
	\item Including recent studies on social media in this area that have not been included in similar efforts.
\end{itemize}

%% file: characteristics.tex
\label{sec:specs}
%\begin{figure*}
%	\begin{subfigure}{0.5\columnwidth}
%		\includegraphics[height=2.5in,left]{disaster-stages.png}
%		\caption{}
%	\end{subfigure}%
%	~ 
	\begin{figure}
%		{1.5\columnwidth}
		%		\centering
		\includegraphics[width=\columnwidth]{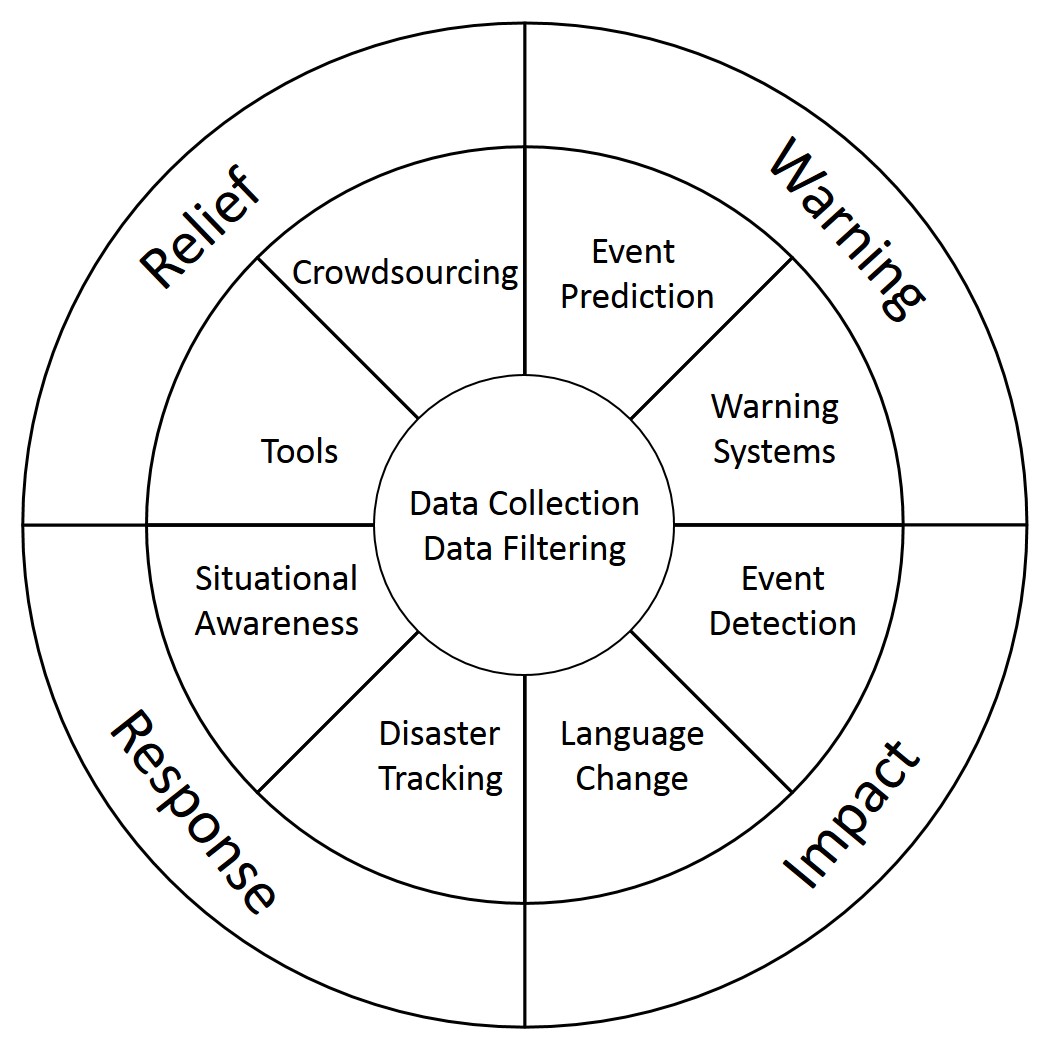}
		\caption{Socio-Temporal stages of disasters which are reflected on social media.}
		\label{fig:process}
	\end{figure}
%	\caption{Socio-temporal stages of disasters. (a) Eight stages of disasters (b) Four stages of disasters which are highly reflected on social media.}	
	%	\centering
	%	\includegraphics[width=\columnwidth]{sandy-wordcloud.png}
	%	\caption{The wordcloud of keywords in tweets related to Hurricane Sandy. The size of each word show how surprising it is.}
	%	\label{fig:sandy-wordcloud}
%\end{figure*}

Lifecycle of disasters consists of several stages. Powell~\cite{powell1954introduction} considers eight socio-temporal stages for disasters: pre-disaster, warning, impact, inventory, rescue, remedy, and recovery. Hill~\cite{hill1962families} also introduces four coarse-grained stages of warning, impact, reorganization, and change. In a model by Office of US Foreign Disaster Assistance~\cite{aim1986schramm}, disaster management lifecycle stages are Hazard Analysis, Vulnerability Analysis, Mitigation \& Prevention, Preparedness, Prediction \& Warning, Response, and Recovery.

Hazard Analysis is concerned with studies on disaster histories and scientific analysis of different disasters. The goal is achieving a thorough understanding of each disaster and how it can affect land, weather, agriculture, and environment. Moreover, malignant effects such as spread of disease, air pollution, and water contamination are studied in this phase. Based on the research in this phase, responders have knowledge about possible outcomes of disasters.

In Vulnerability Analysis the focus is the area and people that are affected by a disaster. Based on the historic record and hazard analysis reports, responders can estimate types and extent of possible damages of a disaster to specific location. Survey and community experience reports can also help with such estimations.

Mitigation and Prevention is about establishment of rules, regulations, and standards that help the community to reduce risks. Land use regulations and building standards are examples of such efforts. To mitigate the risk, organization of relief groups is pre-defined and well-documented. 

Community planning is performed in Preparedness phase. In this stage, communication infrastructure is built and procedures that should be followed after a disaster are defined. Resources, such as food, water, clothes, and medicine, are stockpiled in storages. The community is also prepared by receiving awareness about hazards and actions.

Prediction and Warning is using technology and interpretation methods to forecast disasters and provide early warnings. This phase requires close tracking of disasters and communication with affected areas on the route of the same disaster. Warnings can result in public responses such as evacuation and moving to safe shelters.

Response starts after the disaster onsets. People will be moved to shelters and rescue process for missing persons will start. Responders begin to assess needs of the affected people to make decisions regarding the distribution of resources. Damage to built structures will be estimated and first responders will scatter accordingly.

In the final step, Recovery happens through rehabitation and reconstruction. During this period, affected citizens are in a stable situation and the aim is returning to normal life. Some activities are rebuilding failed structures, providing temporary/permanent housing, reestablishment of agriculture, and securing water sources. 

Although people use social media in all stages of disasters, some stages receive more attention. Volume of data is a major factor in methods that use social media posts. Hence, the phases for witch social media can be used is dependent on how active social media users are during that period of time. The stages that are highly reflected on social media are Warning, Impact, Response, Relief.

Four mentioned stages of disasters are shown in Fig.~\ref{fig:process}. Warning is facilitated by event prediction and warning systems that use social media data. Impact is the time at which disaster hits the area. Onset of disasters can be detected using changes in the behavior of social media users such as language change. In Response phase, occurrence of the event has been confirmed and social media can be extensively used to gain situational awareness and track changes in the status of disaster and affected people. Relief is the stage in which volunteers are engaged to empower tools that facilitate relief efforts. Social media provides a platform to share information and arrange volunteer efforts. In the following sections, we extensively discuss each of the aforementioned stages and how social media can be used to facilitate the efforts of disaster responders.

Valuable insight that is obtained from social media during disasters, in all four stages, is highly dependent on data and its quality. Numerous posts are published in coarse of disasters, however, they are a mixture of informative and non-informative posts. Non-Informative posts can be in form of rumor, spam, bot-generated content, and daily chatter. These posts need to be removed before any anlysis is performed. Hence, the core of Fig.~\ref{fig:process} is data collection and filtering which is required for all the four stages.

%% file: data.tex
Data collection and filtering is the core of disaster management using social media. Algorithm that are used for warning, detecting the impact, relief, and response all depend on the posts which are published on social media and their quality. Two tasks need to be performed in this regard: maximizing the amount of relevant data to the disaster and removing non-informative posts. 

%% file: data_extraction.tex
\label{sec:data_extraction}
% CATEGORIES OF DISASTER RELATED TWEETS EXTRACTION
Disaster-related tweets are extracted using lexicon-based~\cite{imran2013practical,purohit2013emergency} or location-based~\cite{mahmud2012tweet} methods. The former uses a set of keywords that are generated by experts and the latter collects all the tweets that are associated with a specific location. Tweets that are filtered using keywords are only a fraction of all the disaster-related ones~\cite{bruns2012tools} and tweets with location information are quite rare. Hence both methods lack completeness and have low coverage.

\subsubsection{Lexicon Unification and Extension}
One way of increasing the visibility of disaster-related tweets is extending expert-defined keyword sets. For example, ``CrisisLex''~\cite{olteanu2014crisislex} is a a lexicon that increases the portion of disaster-related tweets that are captured from the Twitter Streaming API. The effort is towards finding one set of keywords which are extensively used in different disasters (hurricane, tornado, flood, bombing, and explosion). The process starts with extracting the tweets that contain any word from a set of expert-suggested keywords. These tweets are manually labeled to remove the ones which are not releated to the disaster. From the crisis-related tweets, words and phrases (consisting of two words) that appear in at least 5\% of tweets form CirisLex. Another lexicon is ``EMTerms 1.0''(CrisisLex and EMTerms 1.0 can be obtained from http://crisislex.org/crisis-lexicon.html) that includes more than 7,000 words categorized into 23 groups was introduced. Their method starts with the keywords of four major events and then extend the lexicon using Conditional Random Field (CRF) on another 35 disasters~\cite{temnikova2015emterms}.

Another way is providing instructions for users on how to tweet regarding a disaster. Microsyntaxes, the instructions, unify the format of posts and make it possible for machines to automatically extract all the posts on a specific issue. ``Tweak the tweet''~\cite{starbird2010tweak} is a microsyntax introduced after Red River Floods, 2009. The authors show that visibility of disaster-related tweets increases when users are instructed to use hashtags such as \texttt{\#fargo}, \texttt{\#redriver}, and \texttt{\#flood09}.

\subsubsection{Location Estimation}
\label{location}
% EStimating the location of tweets: 1- content 2- metadata 3- network
To overcome the challenge of location sparsity in social media data, several methods have been proposed to estimate the location of posts or users. Content of posts, activity characteristics, profiles, and networks of users are exploited to estimate the location in which a user is based or the post is originated from. The granularity of estimation differs from one method to another. Some approaches estimate the coordinates, some remain in the city-level, and some only focus on a disaster areas that can be limited to a neighborhood or expand to several cities or states.

% Content
% Linguistic patterns of tweets
Content is frequently used to estimate the origin of posts. N-grams and ``crisis-sensitive'' features such as ``in'' prepositional phrase (such as ``in Boston''), existential ``there'' (which usually describes an abstraction), and part-of-speech tag sequences are signals that discriminate in-region posts from out-region ones in course of a disaster~\cite{morstatter2014finding}.
% Using user profile and information in the tweet
% My COMMENT: they compare the words of tweets from inside and outside of the area at the same time, that's probably why it works. No historical data needed.
Moreover, posts from a disaster area are less likely to include multiple hashtags, action words, and reference entities. Majority of such posts are original and contain URLs~\cite{kumar2014behavior}. 
%Social media posts also contain contextual information which is helpful for location estimation. For example, in-region posts are more likely to be generated using a mobile device~\cite{kumar2014behavior}.
% Content of the tweets for city-level location estimation
% My comment: it cannot be used for disaster time because the language changes during disasters. Historical data needed.
%Similar to disasters which cause in-region posts to contain specific features, posts which are published from the same area (city) tend to have content similarities such as word usage. If a post contains several of highly observed words (n-grams) from an area, it is most likely to be originated from there~\cite{cheng2010you}.
% City-level location for USERs using tweet's text and metadata
Posts from the same location frequently use similar words ~\cite{cheng2010you} and rarely use words that are used in other locations~\cite{han2013stacking}. 

For locating users, the most intuitive features are geolocation or location field in their profile, the location of the websites that they linked to (which can be obtained using the IP or country code), time zone, and UTC24-Offset. These features can be combined using the stacking method~\cite{wolpert1992stacked} by considering an importance weight for each feature to find the most probable location of the user~\cite{schulz2013multi}.

When location-indicating features are not available for a user, their location can be estimated using the location of users surrounding them. 
%Finds the location of users using their physical distance to their friends
Backstrom et al.~\cite{backstrom2010find} observe that there is a power law relation between physical distance and the probability of existing a social link. Based on this finding, they propose a maximum likelihood prediction method that indicates the most probable location for a user given its neighbors. 
%Location estimation based on social "relations"
Based on triadic closure, if user $a$ is connected to users $b$ and $c$, $b$ and $c$ are more likely to be connected to each other~\cite{kossinets2006empirical}. In ``Triadic heuristic''~\cite{jurgens2013s}, the location of users is estimated as the geometric median of their neighbors who are in triadic closure with them. 
%In case that such neighbors do not have a location or the individual has no neighbors in triadic closure, the location would be the geometric median of all its neighbors' locations.
%Li et al.~\cite{li2012towards} combine two signals to predict the location of a user: 
Moreover, users are more likely to follow users nearby and more often mention the location in which they live~\cite{li2012towards}. 
%However, users are different the in the distance in which they link to other users; i.e. celebrities have links to users who live in further distances in comparison to normal users.

%In addition to content, posting behavior of users differs in areas apart. Each time zone can be profiled using the normalized posting frequency of users in each minute of the day. Given the posting times of a user, his most probable location (in timezone granularity) can be found by comparing its posting behaviors with different areas~\cite{mahmud2012tweet}.

% Hybrid method = lexicon + location!
Palen and Anderson~\cite{Palen224} introduce the concept of ``contextual streams'' to combine Lexicon and location in order to overcome the issue of incompleteness. They use a set of broad terms (such as ``frankenstorm'' and ``sandy'' in the case of hurricane Sandy, 2012) to collect the first set of tweets. Then, they find the users who have geolocated tweets from their desired location. Finally, they collect their most recent 3200 tweets and extend their previous dataset. Using this information, they can compare the activities of users who are located on the site of the disaster before, during, and after its occurrence.

%% file: data_filtering.tex
\label{sec:data_filtering}
Filtering social media data is a necessary process before exploiting it in any stages of a disaster. The data which is overwhelmed with unwanted content does not show real opinion of the crowd. Hence, any insight based on this fabricated data is flawed. Unwanted data such as rumors and spam have been studied extensively, however, prevalence of social bots has created new challenges. 
\subsubsection{Rumors}
A rumor is an unverified piece of information that circulates in the situation of uncertainty by people who want to make sense of their situation~\cite{difonzo2007rumor}. Rumors, when verified, form two categories of true and false. False rumors can cause panic and distress especially after the onset of a major disaster. A bold example of false rumors is reported after a few cases of Ebola infection were reported in Newark, Miami Beach, and Washington DC. The rumors state that Ebola can spread through water, food, and air~\cite{luckerson2014fear}. Also after hurricane Sandy, several false rumors have been observed. 
%For a list of rumors refer to the article by Kashmir Hill in Forbes magazine, https://goo.gl/udsdGF}. 
More than ten thousand tweets which contained fake images circulated after hurricane Sandy and the analysis show that top 30 users in this diffusion process were responsible for 90\% of the tweets~\cite{gupta2013faking}. This observation verifies the fact that these images were injected into the stream of Sandy-related tweets by a small group of illegitimate users.
%\footnote{To view a set of images related to hurricane Sandy that are verified as real or fake refer to the article by Alexis C. Madrigal in the Atlantic Magazine, https://goo.gl/pORZ5.}.

There are two mechanisms for manually correcting false rumors after a disaster. The first one is the self-correcting power of the crowd. According to disaster psychologists, social media systems will eventually correct erroneous information as both the sender of the information and other members of that community seek validation of their posts~\cite{crowe2012disasters}. A study by Mendoza et al.~\cite{mendoza2010twitter} shows that majority of tweets which are related to a true rumor are affirming and this percentage is much lower in regard to false rumors. This suggests that Twitter community works as a collaborative filter for information. However, this mechanism is effective when a false rumor is not originated intentionally and enough time is given to this process to converge. In the aftermath of disasters such as hurricane Sandy, these conditions are not satisfied and so there is a need for emergency managers and officials to intervene which is the second method of correction. An example of this mechanism is the tweet which was posted by ConEdison energy company indicating that none of its employees have been trapped in a damaged plant and end the circulation of rumors about it (this tweet can be accessed via https://goo.gl/TC1Slx).

The aforementioned rumor correction methods are time-consuming and dependent on manual efforts which make them less effective as the social media becomes more prominent as the information channel after disasters. More than 300,000 tweets per minute after Japan Earthquake in 2011 and more than 20 million tweets after hurricane Sandy show the volume of data that needs to be processed to extract rumors. To overcome this challenge, methods have been proposed to automatically detect tweets in the swarm of social media posts by using their specific behaviors. For example, the diffusion process of rumors is different from those of normal posts. The originality of posts is lower (most of the posts are retweets).  The number of users involved in the cascade is low in comparison to its virality. Also, the depth of the resulting cascades are lower than those of normal posts as users receive information by search and the posts, mainly, do not diffuse in the friendship network~\cite{gupta2013faking}.

\subsubsection{Spam}
\begin{figure}
	\centering
	\includegraphics[width=\columnwidth]{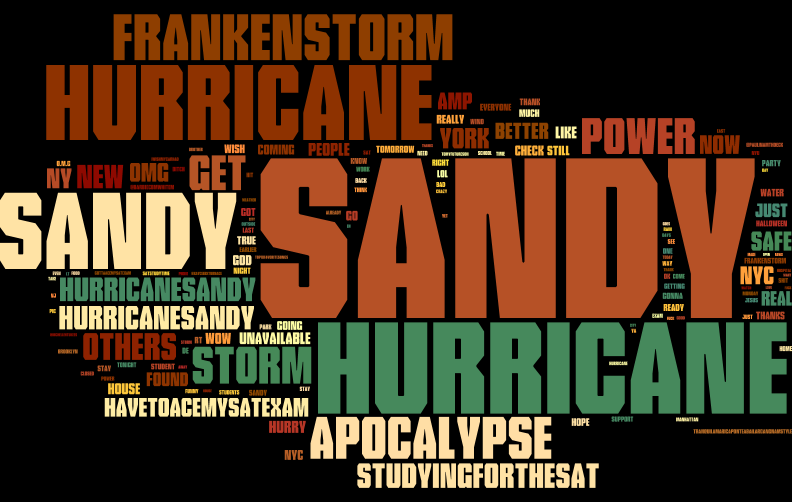}
	\caption{The wordcloud of keywords in tweets related to Hurricane Sandy. The size of each word corresponds to its surprise level in the area under study.}
	\label{fig:sandy-wordcloud}
\end{figure}

Spam is the ``content designed to mislead or content that the site’s legitimate users do not wish to receive~\cite{heymann2007fighting}.'' Spammers degrade the credibility of social media and are capable of deviating the discussions and affect the popularity of topics. An example of such manipulations is promoting the material for Scholastic Assessment Test (SAT) that was happening close to the occurrence of Hurricane Sandy. The wordcloud (Figure \ref{fig:sandy-wordcloud}) that has been generated using the surprise level of words~\cite{sampson2015real} from Twitter data after this disaster shows \texttt{\#HaveToAceMySATExam} as one of the keywords that stands out among thousands of others. 

Social media sites such as Twitter and Facebook actively remove who violates their regulations and specifically spammers. The dominance of spammers becomes clearer when Twitter suspension included 50\% of the accounts that were created in 2014~\cite{koh2014only}. Spammers, although elaborate, show specific characteristics that discriminate them from legitimate users. Posting numerous posts on a handful (and even just one) topics throughout the lifetime of an account is a sign of spamming as marketing spammers advertise a specific product and publish many posts to gain visibility. Profile of spammers in most cases does not show their age and gender~\cite{zafarani201510}. Their connections are different from normal users as the majority of their connections are not reciprocal~\cite{wu2015social}.

\subsubsection{Bot Generated Content}
A malicious bot is a hijacked or adversary-owned account which is controlled by a piece of software. Bots, that can be automatically generated in large numbers, are overwhelming social media and leave major tracks. In the 2010 Massachusetts Senate Election, a candidate gained 60,000 fake followers on social media by exploiting bots. Such activities manipulate the opinion of the crowd by promoting a specific topic or supporting a specific figure. In these cases, trending topics and popular users are not necessarily real. 

Three major methods are proposed to collect bots to observe and study. The prominent method is manual annotation which is expensive, time-consuming, and error-prone~\cite{chu2010tweeting,ratkiewicz2011truthy,xie2008spamming}. The second one is using the suspension mechanism of social media sites such as Twitter. There is no explicit cost associated with this method, however, it is time consuming and bot behavior is one reason for suspension. The process is sampling users, waiting for a period of time, and then re-examining the status of the sampled user. The ones that have been suspended in that period of time would be considered as bots~\cite{john2009studying,thomas2012adapting}. In the third method, a set of bots (honeypots) are created to lure other bots in the wild to interact with them. Honeypots are controlled by the researcher to tempt specific bots and avoid interference with the activities of normal users~\cite{lee2011seven,morstatter2016new}.

%% file: warning.tex
\label{sec:warning}
Accurate and timely warnings could prevent death rates by providing the time that is critical for evacuating elevators or halting medical operations. Warning systems have improved drastically in recent years but they are not perfect yet. In a recent case, Hurricane Matthew was reported as a category 4 hurricane as it approached the Florida coast~\cite{palm2016hurricane}, but it had one official U.S. landfall on the southeast of McClellanville, South Carolina as a category 1 hurricane~\cite{url2016hurricane}. Social media can be used as a complementary source of information to improve predicting events and providing warnings.

\subsection{Event Prediction}
Social media has been used for predicting events that will happen in near future. Forecasting the popularity of products, movie box-office, election results, and trends in stock markets are examples of such predictions~\cite{yu2012survey}. 
%Social media posts have been also used for predicting disasters such as earthquakes and protests.

Prediction is based on features of social media posts. Increase in the number of posts which are related to a topic~\cite{asur2010predicting} can be indicator of its future popularity. Changes in the patterns of using specific words in a area shows onset of an event~\cite{sampson2015real}. Also, sentiment of posts can show future status of a product~\cite{mishne2006predicting}. Crime prediction is also possible by semantic role labeling which is used for both finding the events and entities involved in them~\cite{wang2012automatic}.

Prediction method based on the extracted features can vary based on the problem. Regression method have been used for prediction popularity of posts~\cite{szabo2010predicting} but do not perform well when sentiment data is being used~\cite{zhang2009improving}. For predicting election results, Tumasjan et al.~\cite{tumasjan2010predicting} use number of tweets mentioning political parties and their sentiments as indicators of popularity and political views toward them~\cite{tumasjan2010predicting}.

There is no prediction method with perfect accuracy. However, early detection of natural disasters reduces hazards in nearby locations. For example, quakes in areas with geographic proximity are used to predict earthquakes seconds before they happen~\cite{faulkner2011next}. This process has been used to build warning systems which will be explained in Section \ref{sec:warning}.

%???? Write about some papers from~\cite{yu2012survey}.
\subsection{Warning Systems}
``Warning systems detect impending disaster, give that information to people at risk, and enable those in danger to make decisions and take action''~\cite{sorensen2000hazard}. There has been a significant improvement in forecasting and warning systems especially for hurricane and earthquakes. Meteorologists can now forecast a hurricane 2 to 6 days before it hits an area and Global Seismic Network constantly monitors activity bellow Earth's surface. However, lack of complete data on natural hazards, monitoring instruments, and high dynamic nature of them keep accurate forecasting and warning a challenge~\cite{reese2016how} and ``a 100\% reliable warning system does not exist for any hazard~\cite{sorensen2000hazard}''. 

%Warning systems for disasters such as earthquakes have advanced drastically by ~\cite{reese2016how}. However, there are still dead zones such as the ocean floor and predicting exact time that an earthquake happens is still not feasible. As opposed to earthquakes which have been thoroughly studied and are constantly monitored, most vulnerable areas to landslides and landslide hazard maps are far from being complete. , however, the intensity of them cannot be predicted. In a recent case, Hurricane Matthew was reported as a category 4 hurricane as it approached the Florida coast~\footnote{\texttt{https://goo.gl/ZW33U3, accessed 10 October 2016}}, but it had one official U.S. landfall on the southeast of McClellanville, South Carolina as a category 1 hurricane~\footnote{\texttt{https://goo.gl/8Rfn6S, accessed 10 October 2016}}. 

%Also, there is no global monitoring system for volcanoes. 
%Warning systems have been used for more than 35 years in disasters such as hurricane, flood, fire, earthquake, and nuclear power. . In the case of an earthquake, shaking in the locations further from the origin of an earthquake happen a few seconds after the accelerometers capture the initial shake. This time is critical for tasks such as evacuating elevators or halting medical operations. Currently, United States National Geographic Survey (USGS) monitors more than 2,000 real-time earthquake sensors. However, these sensors are not perfectly accurate and using other information sources can be used as a secondary check.

Social media facilitates is also used to deliver official and non-official warnings. Emergency managers and governmental organization post their warning messages via social media to be broadly accessed by the public~\cite{houston2015social}. Citizens also report warnings and advice about possible hazards~\cite{imran2013extracting}.

One source of information that can be used to improve accuracy of warnings is data of built-in accelerometers in cell phones. This data can be used for quick detection of earthquakes and estimating their intensity and effect. The measurements by these sensors which are transmitted before the loss of communication are used for estimating the degree of damage to different areas; the task that can take up to an hour when performed by helicopters~\cite{faulkner2014community,faulkner2011next}.

Social media is another source of information for warning systems. USGS uses tweets to check the accuracy of sensor reports and faster detection of earthquakes.  Disasters such as Sichuan earthquake in 2008 show that Twitter is faster at reporting earthquakes than USGS. Earthquakes can be detected using tweets by 60 seconds earlier than sensors which is a valuable time for warning areas under danger and start evacuation~\cite{Ellis2015Twitter}. In another effort, Sakaki et al.~\cite{sakaki2010earthquake} consider each Twitter user as a sensor. The tweets by these sensors will be used to detect the occurrence of disasters and estimate their location.
% The same news as the above footnote: http://money.cnn.com/2015/10/07/technology/twitter-earthquakes/
%For this task, the same event detection methods which are used for earthquake sensors is used for tweets. USGS scientists have observed that the tweets which contain first-hand reports are mostly less than 7 words and do not contain URLs.
%Although major advances in hazard warning systems have been achieved, ``a 100\% reliable warning system does not exist for any hazard~\cite{sorensen2000hazard}'' and major disasters will result in casualty and damage to infrastructures in most cases and relief delivery and rebuild are necessary for impact, reorganization, and change stages of such disasters.

%% file: impact.tex
\label{sec:impact}
Events are widely reflected on social media even before they are reported by news agencies and official sources. For example, in London Subway Bumbing and Virginia Tech Shooting, social media has been the primary source of information. As the event happens, social media posts show anomalies which can be captured by event detection methods. One of the major impact of disasters on social media are changes in the language of posts.

\subsection{Language Change}
\label{sec:lang_change}
Qualitative studies on social media show that language of users change after disasters due to distress. 
%The distress which is caused by disasters is reflected in the behaviors of the affected and distant people. In the physical world, they check the safety of their loved ones and make donations and on social media the language they use differs from normal situations. This language change has been qualitatively and quantitatively studied. 
A study~\cite{cohn2004linguistic} on LiveJournal after 11 September 2001 shows that emotional positivity decreased and cognitive processing (intellectually understanding the issues), social orientation (how much other people are mentioned), and psychological distancing increased after the attack. Emotions of the crowd after disasters is another area which can be tracked and used by means of sentiment analysis tools. The sentiment of users in their posts can help distinguish the posts that come from the affected area and track emotional situation of people in different stages of disasters~\cite{beigi2016overview}. 

In quantitative analysis of language change after disasters, language is statistically modeled at the level of sentences or topics. As the disaster happens, the language of the affected people on social media changes. Several measures have been developed to quantify language change. Here we introduce some of the measures that have been used in event detection methods. These methods are based on the assumption that when the language of the people in a specific area changes more than a threshold, it is a sign of an irregular event in that region (for surveys on other categories of event detection refer to~\cite{atefeh2015survey} and \cite{cordeiro2016online}). Here, we enumerate some of these measures in sentence-level and topic-level language models.

\subsubsection{Sentence-Level Language Change}
Sentence-Level language change measures the novelty of a sentence in comparison to a presumed set of sentences. For event detection in a region using Twitter, language model of a tweet is compared to the language model of the tweets that have been posted in normal situation from the same region. Kullback-Leibler (KL) divergence~\cite{kullback1951information} is a measure that calculates divergence between two sentence-level language models, $\Theta_1$ and $\Theta_2$ as defined in Equation \ref{eq-KLDivergence}.
\begin{equation}
\label{eq-KLDivergence}
KL(\Theta_1||\Theta_2) = \sum_{w}p(w|\Theta_1)log\frac{p(w|\Theta_1)}{p(w|\Theta_2)}
\end{equation}

A sentence-level language model is a statistical model of sequences of words (i.e. sentences). As shown in Equation~\ref{eq_sentence_prob}, the probability of observing a sentence, $w_1w_2...w_n$, is calculated based on the assumption that probability of observing each of its words, $w_i$, is dependent on the previous words.
\begin{equation}
\label{eq_sentence_prob}
\begin{split}
p(w_1...w_n) &= p(w_1)\times p(w_2|w_1)\times \\
& p(w_3|w_1w_2) \times...\times p(w_n|w_1...w_{n-1})
\end{split}
\end{equation}

A common method for calculating the probabilities in Equation \ref{eq_sentence_prob} is Maximum Likelihood estimation. In this method, the probability of observing word $w_n$ after observing the sequence of $w_1...w_{n-1}$, $p(w_n|w_1...w_{n-1})$, is calculated using the conditional probability in Equation~\ref{eq_word_prob}.
\begin{equation}
\label{eq_word_prob}
p(w_n|w_1...w_{n-1}) = \frac{C(w_1...w_n)}{\sum_{W}C(Ww_n)}
\end{equation}
Where $C(w_i...w_j)$ is the frequency of observing the sequence of words $w_i$ to $w_j$ and $W$ is any possible sequence of $n-1$ words in the corpus.
%Formally, in a sequence of $n$ words, $w_1w_2...w_n$, probability of observing word $w_n$ based on the previously observed $n-1$ words, $w_1w_2...w_{n-1}$ is calculated using $P(w_n|w_1w_2...w_{n-1})$. 

One challenge of using Equation~\ref{eq_word_prob} is calculating the denominator. Computing the denominator is computationally expensive in large corpora due to a large number of possible sentences. To overcome this challenge, $n$ is usually set to 1, 2, or 3 which yields to unigram, bigram, or trigram language models respectively~\cite{jurafsky2014speech}.

Another challenge is calculating the probability for the sequences (sentences) that have not been observed in the corpus. To overcome this problem, interpolation techniques can be used in which the probability of observing a sequence is calculated by using the probabilities of shorter sequences. For example, in a trigram language model, probability of observing a trigram can be calculated by mixing probabilities of bigrams and unigrams as shown in Equation~\ref{eq_interpolation}.
\begin{equation}
\label{eq_interpolation}
\begin{split}
\hat{p}(w_n|w_{n-2}w_{n-1}) &= \lambda_1p(w_n|w_{n-2}w_{n-1}) + \\
&\lambda_2p(w_n|w_{n-1}) + \lambda_3p(w_n)
\end{split}
%\hat{p}(w_n|w_{n-2}w_{n-1}) = \lambda_1p(w_n|w_{n-2}w_{n-1}) + \\\lambda_2p(w_n|w_{n-1}) + \lambda_3p(w_n)
\end{equation}

\subsubsection{Topic-Level Language Change}
A set of event detection methods are based on the assumption that burst in observing explicit topics is a sign of the occurrence of an event. Explicit topics are the ones assigned by the author of posts and are mostly known as hashtags in social media. Examples of hashtags are \texttt{\#frankenstorm} and \texttt{\#hurricane} in the case of Hurricane Sandy (2012). The burst is considered as an unexpected rise in the frequency or a transformation of hashtags~\cite{xie2013topicsketch}. Each hashtag can be represented in time-frequency space using continuous wavelet transformation on its frequency. Wavelet peaks show unusual bursts in observing that hashtag~\cite{cordeiro2012twitter}.

Occurrence of events can also be detected using hidden topics. In these methods, language is modeled as a probability distribution over topics. The assumption is that there is a fixed set of hidden topics in a corpus, each document is a random mixture of these topics, and each topic is a distribution over words. Using Latent Dirichlet Allocation (LDA)~\cite{blei2003latent}, we can extract the probability of each topic in a document. Moreover, word distribution in each topic gives an insight on what the topic is about. To measure language change in course of a disaster, posts (such as tweets) are considered as sentences. If hidden topics of these posts largely deviate from topics of posts in regular situations, it will be considered the signal of an event.

Event detection methods based on hidden topics detect abnormal topics in a specific region. Chae et al.~\cite{chae2014public} have extracted major topics in the area of interest using LDA. A time series based on the daily message count on each topic is generated Seasonal-Trend Decomposition Procedure Based on Loess (STL)~\cite{cleveland1990stl} has been used to decompose the time series into three components: a trend component, a seasonal component, and a remainder. The remainder is supposed to be identically distributed Gaussian white noise. However, when the remainder has a large value, it indicates substantial variation in the time series. This variation can be considered as novelty or abnormality in the language. In this work, if the seven-day moving average of the remainder values has z-score higher than 2, events can be considered abnormal in 95\% confidence.

\subsection{Event Detection Methods}
\label{sec:event_detection}
%???? Follow~\cite{atefeh2015survey,cordeiro2016online} and refer to main categories of event detection on social media.
Events are real-world occurrences that unfold over space and time and. The goal of event detection methods is extracting events in a stream of news or social media posts~\cite{allan1998topic}. Event detection using social media has been extensively studied and the different categorizations are available for proposed methods in this area.

When there is no information about future events available, the event detection method falls into the unspecified category. In this category, detection methods are based on bursts or trends in the stream of posts~\cite{popescu2010detecting}. In the specified event detection methods, contextual information such as time and venue are available for the anticipated event~\cite{becker2011automatic}.

Another categorization of events is new versus retrospective. New event detection is extracting previously unseen events from a stream of posts as they come. Retrospective event detection also finds unseen events but the data source is an accumulation of historic posts~\cite{allan1998topic}. Clustering methods are the most common in detecting both types of events~\cite{atefeh2015survey}. But there are also supervised methods such as Naive Bayes~\cite{becker2011beyond} and gradient boosted decision trees that have been used for new event detection~\cite{popescu2011extracting}.

Clustering methods focus on documents and grouping them based on similarities, i.e. they are document-pivot. There is a group of feature-pivot techniques that use changes and bursts in features of documents to detect events. These features include frequency of specific keywords~\cite{power2014emergency}, surprise level of relevant keywords~\cite{sampson2015real}, and statistical features of posts (i.e. word frequencies)~\cite{sakaki2010earthquake}.

%% file: response.tex
\label{sec:response}
In the chaotic environment of disasters, emergency responders want to acquire a big picture of the event and actionable insight~\cite{castillo2016big}. Preliminary assessment of the disaster such as the area which is affected, the number of casualties, and failed infrastructures are obtained in the ``big picture''. ``Actionable insights'' are concerned with specific information with more details such as requests for help. 

%The first step in acquiring the big picture is collecting relevant data which comes from the affected people (Section \ref{sec:extraction}). This data needs to be cleaned by removing rumors, spam, and bot-generate content. The cleaned data is more accurate in reflecting real status of the disaster (Section \ref{sec:data_filtering}). The big picture of the event can be in part achieved using the topics which are discussed by people and trends which emerge and fade throughout the disaster (Section \ref{sec:tracking}). Finally, actionable insight is gained by extracting social media posts which contain situational awareness pieces of information (Section \ref{sec:situational_awareness}).

%% file: tracking.tex
\label{sec:tracking}
Systems that monitor social media for crisis-related purposes use computational capabilities such as collecting data, Natural Language Processing, information extraction, monitoring changes in data statistics, clustering similar messages, and automatic translation~\cite{imran2015processing}. Three important results of these computations are topics, trends, and memes. In the remaining of this section, we discuss how to gain and use these three for tracking the status of disasters.
\subsubsection{Topic Discovery and Evolution}
% LDA and semantic based topic modeling ??????????????????????
%They study how tweets on different topics differ in persistency~\cite{romero2011differences}. ???????????????
Several methods are used for discovering topics from a corpus of text (news articles, tweets, Facebook posts, etc): Latent Dirichlet Allocation (LDA), Non-negative Matrix Factorization (NMF), Term frequency-inverse document frequency (tf-idf), and PLSI. LDA~\cite{blei2003latent} considers a fixed number of latent topics for the corpus and finds the probability of each document belonging to each topic. A topic itself is a distribution over words (vocabulary). In NMF~\cite{lee2001algorithms}, the document-word matrix would be factorized into two matrices, document-topic and topic-word. NMF describes both documents and terms in the environment of latent topics. tf-idf is widely used in Information Retrieval. Using tf-idf, each document is represented using a vector of size $V$ (vocabulary size). Element $ij$ of this the tf-idf vector is the frequency of word $i$ in document $j$ time the inverse of the total frequency of word $i$ in the whole corpus of documents. Probabilistic latent semantic indexing (PLSI) is based on the assumption that each document has one topic and words and documents are conditionally independent given the topic of the document. Hence the probability of a word $w_n$ in document $d$ is calculated using $p(w_n,d) = p(d)\sum_{z}^{} p(w_n|z)p(z|d)$.

% Topic Evolution
As the disaster proceeds in its stages, concerns of the affected people evolve and hence the topics of discussion. There is a body of research modeling evolving and fading topics which are usually known as topic discovery and evolution (TDE) methods. Vaca et al.~\cite{Vaca2014time} and Kalyanam et al.~\cite{kalyanam2016event} in separate works, similarly model this problem using a modified version of NMF. In these proposed methods, besides decomposing the document-word matrix into document-topic and topic-word matrices, a matrix M would be considered which models how topics evolve from each time step to the next. Hence one topic matrix is calculated at each point of time. The entropy of topics in each time step indicates its status, continuing, evolving, or new. In another work~\cite{kalyanam2015leveraging}, besides using the evolution of topics, they have also exploited social context. Their method is based on the assumption that members of one community have similar interest in the topics and show that this information improves topic discovery results especially when the topic has a large set of keywords associated with it or evolves much over time and hence is difficult to detect. However, the persistence of the user who shows interest in these topics help the detection methods.

\subsubsection{Trend Mining}
% What is the limit for finding trending topics
Mathioudakis et al.~\cite{mathioudakis2010twittermonitor} define trends as ``set of bursty keywords that occur frequently together'' which are driven by events and breaking news. Naaman et al.~\cite{naaman2011hip} define a score for each word on Twitter and the top 30 words will be considered as trending in each hour. The score is calculated using \textit{((word frequency in a specific hour)-(average frequency of the word in this specific hour across weeks))/(standard deviation of the frequency of this word in this hour across weeks)}.
%Classification of trends based on the frequency of posts on that trend over time
Trends on Twitter are either caused by an external happening (such as a natural disaster) or are specific to the social media (a tweet by a famous user). Based on the source, they are divided into exogenous and endogenous~\cite{naaman2011hip}. 

% Detecting new trends on Twitter and summarizing them "without tracking".
Trends indicate what are the major subjects that people talk about. Resources that the affected people need and the issued that they talk about can be a subset of trends after a disaster. TwitterMonitor~\cite{mathioudakis2010twittermonitor} is a system that detect new trends/topics by finding bursty keywords and clustering them based on co-occurrence. In the analysis of trends, they summarize each trend using the most frequent keywords, other words which are not as dominant but they are highly correlated to the frequent keywords, and also the named entities and sources (URLs) which are mentioned frequently.

% Visualizing topic evolution (birth, death, merge, and split) using river flows
Cui et al.~\cite{cui2011textflow} propose a clustering-based model for visualizing topic discovery and evolution. They use a heuristic to hash the topics in each time step and for discovering new topics and detecting the death of topics (these two are critical topics) they compare the hashes from one time step to the next. They also monitor the merge and split of topics and these changes in the topics are shown in a river flow visualization. Width of a flow shows the occurrence number of all involved keywords in that topics and the merge, split, birth, and death of topics are shown by colors.
%, they measure the words in one cluster at the current time to other clusters in the previous time step and the percentage of overlap shows if a topic is a split from another topic or two topics have merged with each other from the previous time step. A
%These two categories are further classified as bellow:
%\begin{itemize}
%	\item Exogenous trends
%	\begin{itemize}
%		\item Broadcast-media events
%		\item Global news events
%		\item National holidays and memorial days
%		\item Local participatory and physical events
%	\end{itemize}
%	\item Endogenous trends
%	\begin{itemize}
%		\item Memes
%		\item Retweets
%		\item Fan community activities
%	\end{itemize}
%\end{itemize}

Another approach is a fine-grain classification of trends that is based on comparing the frequency of words/hashtags before and after a spike. Lehman et al.~\cite{lehmann2012dynamical} classify the trends on Twitter into three categories based on the shape of the spike: activity centered before and during the peak, concentrated during and after the peak, and symmetric activity.

% Predicting flu trends for the next week (the percentage of people who will have flu in the next week)
% I doubt that this simple model can predict peaks well and their data has no sudden peaks, I wonder if it can work better than Google!
%Achrekar et al.~\cite{achrekar2011predicting} use a combination of physician visits and the number of flu-related tweets in one week to predict the percentage of flu cases in the next week. Their experiments in 20 weeks in 2009 and 2010 show that they have less than one percent error in their predictions for most of the weeks.

\subsubsection{Meme Tracking}
%Trend Mining/Meme Tracking/Topic Evolution
% When does a topic become a trend or meme???????????????????????????
% How topics evolve over time?????????????????????????????
%In some research, they consider quotes in news articles as memes that evolve and diffuse over time.
%Tracking how the strength of topics in a corpus of documents changes and visualizing it using a river shape.
Memes are short units of text that act as the signature of a topic~\cite{leskovec2009meme}. Harve et al.~\cite{havre2000themeriver} introduce a visualization for thematic change in over time for a corpus of documents. Their method shows a ``river'' of themes that includes colored ``currents''. Each current is a theme and its width at a specific point of time shows its strength. When a current becomes wider, the topic or set of topics associated with it are more dominant in the corpus and as the color changes, the themes in the corpus are is changing.

%Tracking memes and observe how they evolve (Meme can be considered as a topic and this paper as topic evolution!)
% In the second paper by the same group they analyze bias using quotes and how different new media use different quotes from the same speech and 
% how the sentiment and negative words are used by one party when quoting the president's speech.
% A meme is a phrase which gets smaller and different (by changing one word in each step) so the last node in a cluster of nodes may be part of the original phrase. visualization is somehow similar to havre2000themeriver.
Leskovec et al.~\cite{leskovec2009meme} detect and track memes (quotes) in news articles. They generate a graph in which each node is a meme and there is an edge from node $i$ to $j$ if the meme in node $i$ is strictly shorter that $j$ and the directed edit distance to $j$ is less than one. This directed acyclic graph is then partitioned in a way that all the nodes in one cluster can be considered ``belonging'' either to a single long phrase or to a single collection of phrases. To analyze the evolution of memes in a corpus of news articles, they use variations of a meme and extract all the articles that contain those memes and graphical visualize the changes in the volume of corresponding documents over time using stacked plots.

The same concept of memes has been used to monitor bias in news outlet by different parties in the United States. Niculae et al.~\cite{niculae2015quotus} build a bipartite graph from quotes of the president's speeches to news outlets which are used in a matrix factorization method to predict the future quotes of a news outlet based on its previous quotes. Moreover, they analyze the sentiment and negativity of quotes in different outlets and present the bias present in reporting parts of the speeches with a specific choice of negative words and negative sentiment in one of the parties.

%% file: situational_awareness.tex
\label{sec:situational_awareness}
% DEFINITION OF INFORMATIVE TWEETS AND SITUATIONAL AWARENESS
%``Situational awareness is the accessibility of a comprehensive and coherent situation representation which is continuously being updated in accordance with the results of recurrent situation assessments''~\cite{sarter1991situation}. Similar to previous studies~\cite{imran2013practical,vieweg2012situational}, we consider ~\cite{imran2013practical,vieweg2012situational}
Informative posts provide ``tactical, actionable information that can aid people in making decisions, advise others on how to obtain specific information from various sources, or offer immediate post-impact help to those affected by the mass emergency''~\cite{vieweg2010microblogging}. Informative posts can be categorized into six groups~\cite{imran2013practical}: caution or advice, information source (photos and videos), people (missing or found), casualties and damage (infrastructures, injured, or dead), donations (requests and offers for money/good/services), people (celebrities and authorities).
%These categories are further divided into 32 groups using qualitative analysis on Red River Flood (2009), Oklahoma Fires (2009), and Haiti Earthquake (2010)~\cite{vieweg2012situational}. 
%Informative tweets communicate information about social, built, or physical environments.

% REQUESTS AND OFFERS, EXTRACTION AND MATCHING
One attempt toward increasing situational awareness based on social media posts is extracting requests and responses. Varga et al.~\cite{varga2013aid} use content of tweets to extract problems and aid reports on Twitter after Japan earthquake (2010). They use the notion of ``problem nucleus'' and ``aid nucleus'' and exploit features such as trouble expressions (a manually created list of trouble expressions), excitation polarity (excitatory, inhibitory, and neutral), word sentiment polarities and word semantic word class (clusters of words such as food and disease), and location mentions to train a classifier that labels tweets as a problem or air report. 
%For matching the corresponding problem and aid reports, they have another classifier which benefits features such as opposite excitation polarities in problems and matching aids and common semantic word classes.

Another attempt is toward matching requests with appropriate responses. Purohit et al.~\cite{purohit2013emergency} propose a solution using a two-step model. In the first step, they use a binary classifier that uses n-grams and regular expressions to label tweets as requests and offers. In the next step, based on the cosine similarity of tf-idf term vectors, they match requests and offers. In this work they focus on donation related tweets and they consider money, medical, volunteer, clothing, food, and shelter requests/offers as subcategories. Their studies on Hurricane Sandy dataset (2012) shows that the majority of donations lay in ``money'' category.

%% file: relief.tex
\label{sec:relief}
Volunteers are significantly important in the relief process. They post information such as road status and damages to built structures which increases awareness. They also provide technical support such as translation of posts and geotagging them. There are also several systems that are built to exploit and organize such efforts.
\subsection{Crowdsourcing}
\label{sec:crowd_sourcing}
Volunteering is part of how community reacts to disasters~\cite{dynes1970organized} and this process has been facilitated by social media in recent years. Volunteers provide information and resources to the affected people. Example of such efforts are providing temporary housing for stranded people in the US after terrorist attacks in Paris (2015)~\cite{sederholm2015stranded} and offering food and shelter after Hurricane Sandy (2012)~\cite{kanver2012hurricane}.

Digital technologies have broaden domain of volunteer activities in course of disasters. ``Digital volunteers''~\cite{castillo2016big}, either located in the disaster area or in distant locations, ease the relief efforts by providing variate of services. Translation of posts, geotagging posts that indicate an incident on the map, creating maps of open and blocked roads, increasing accuracy of maps by marking built entities on the maps are example of such efforts~\cite{meier2012crisis}.

Many systems have been developed to benefit and organize volunteers who use social media. OpenStreetMap is one of the systems that allows volunteers contribute in generating open source maps by marking entities such as roads and buildings. These maps have been used for disasters such as Haiti earthquake (2010)~\cite{zook2010volunteered}. The details about more systems is presented in Section \ref{sec:tools}.
\subsection{Relief Tools}
\label{sec:tools}
%Volunteering is part of how communities react to disasters. This reaction has been facilitated by and new ways of collaboration have emerged through social media. Millions of posts are published related to major disasters which provide valuable insights and contribute to situational awareness. 
Social media is a unique platform for collaboration between remote volunteers. These volunteers provide technical services such as translation, geolocating posts on the map, and generating maps of the affected area. Several tools have been developed to use crowdsourcing and social media for facilitating volunteering actions.
\subsection{Ushahidi}
Ushahidi is the first large-scale crowdsourcing system for disaster relief. It has been initially developed to map the reports of Kenyan post-election violence in 2008 and since then has been used in many major disasters such as Hurricane Sandy and Haiti Earthquake. Ushahidi is an open source and free systems which can either be deployed on external servers or on Ushahidi's hosting system CrowdMap. When technical knowledge or hosting servers are not available, CrowdMap is a more suitable.

Ushahidi has three main sections: data collection, visualization, and filtering. As the first step, disaster-related data is collected from several sources, web, Twitter, RSS feeds, emails, SMS, and manual comma separated files. The user-contributed information is then visualized on the map. Each point on the map shows one report and when a user zooms out, aggregated number of reports in each area is represented. As the last step, Ushahidi allows users to filter reports based on their types, e.g. supplies or shelter.
\subsection{AIDR}
Artificial Intelligence for Disaster Response (AIDR) is a free software platform which can be either run as a web application or created as its own instance. This system allows the detection of different categories of tweets based on a small sample of labeled tweets. The process has three steps, data collection, annotation, and classification. Tweets are collected based on a pre-selected set of keywords. A small portion of theses tweets is then labeled by volunteers as in-category or out-category. In each disaster, different categories can be considered such as status update, shelter, or food. Labeled tweets which can be as few as 200, will be used as the training set of a classifier which labels remaining set of tweets which were collected based on the keywords. In the training process, n-grams of tweets are used as features and hence the classifier needs to be retrained for every new category and disaster.
\subsection{TweetTracker}
TweetTracker is a system for tracking, analyzing, and understanding tweets related to a specific topic. To track the status of and event, data can be collected using a set of criteria including keywords, location, and users. The source of the data can be chosen from Twitter, Facebook, YouTube, VK, and Instagram. Changes in the total number of post or frequency of posts with specific words can be plotted for different time periods. Moreover, keywords, hashtags, links, images, and videos with their frequencies are available to the user. To better understand the geographic distribution of posts on the globe, the posts which are geotagged will be shown on a map.

\begin{figure*}
	\centering
	\includegraphics[width=\textwidth]{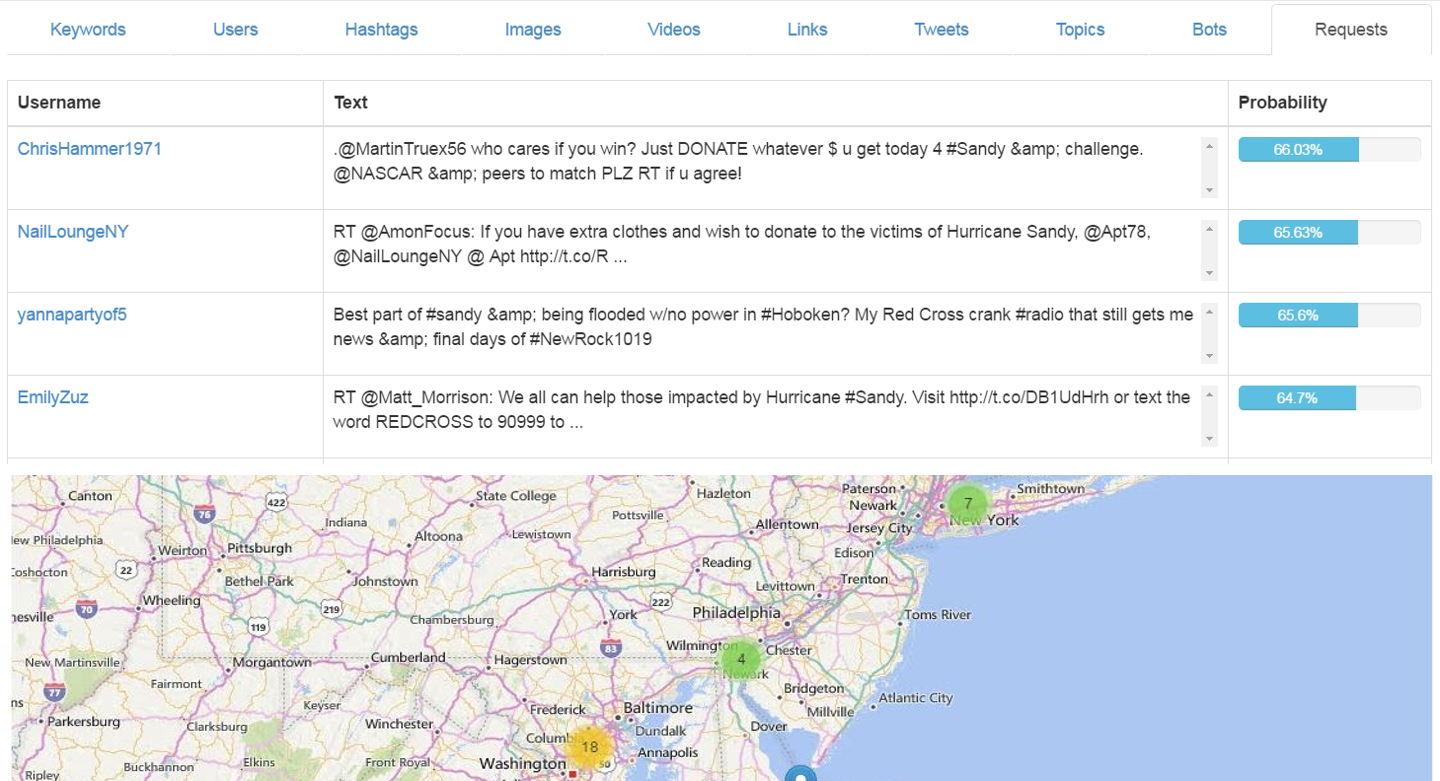}
	\caption{A view of a TweetTracker module which shows requests-for-help tweets related to Hurricane Sandy.}
	\label{fig:request-for-help} 
\end{figure*}

All the features mentioned above are useful for any topic which is discussed on social media. However, there is a module in TweetTracker which is specifically designed for disaster relief. In this module, as the tweets related to the target disaster are captured by the system, the ones which are most probable to contain a request for help will be detected. These tweets in the majority are posted by the affected people and need urgent attention. The classifier for this task works based on both content (n-grams) and metadata of tweets. This brings the flexibility which lets the classifier be used for different disasters. The more certain requests that have geolocation will be also shown on the map. A view of this system is shown in Fig.~\ref{fig:request-for-help}.
\vspace{5mm}

%% file: conclusion.tex
Disasters are widely reflected on social media and this swarm of information provides valuable insight for governments, NGOs, emergency managers, and first responders. It also helps the affected people keep in touch with their loved ones, finds information about the status of the disaster, and be informed about emergency contacts. Social media is the new way of communication in emergencies which transfers information before and faster than traditional news media. It is prevalent in such a way that disaster responders encourage citizens to exploit it to take some load off the cellular systems which usually becomes overwhelmed by calls and text messages in chaotic situations. On social media, both responders and affected people can broadcast information and in contrast with traditional media, people can also provide feedback to officials. 

These potentials have encouraged responders to benefit social media in large extent in recent years. However, there are challenges associated with this task. Social media posts come at a fast pace and immense volume. Moreover, it is challenging to collect all the posts which are related to a disaster due to the restrictions posed by social media websites. The collected data contains daily chatters, prayers, and opinions and is only in part insightful information which adds to the situational awareness. Another issue is malicious content such as spam and rumors which can cause panic and stress, especially when produced in large scale using bots. Even after the data is filtered from all the aforementioned posts, it is still challenging to extract specific groups of information such as requests for food or shelter.

In this paper, we focused on four phases of disasters: warning, impact, response, and relief. These stages are the ones during which computer science has been helpful the most. For each stage, we introduced some of the recent impactful research and response has the greatest portion because most of the social media activity after disasters are in this phase. Social media posts can be used to detect the onset of events even prior to official sources and be used in warning systems. Several methods have been used to increase the coverage of data which is collected regarding disasters and filtering it from unwanted content. We also mentioned tools, such as Ushahidi, AIDR, and TweetTracker, for exploiting volunteer efforts.

%% file: future.tex
Although disaster management has achieved major advancements in using social media, there are still several challenges to overcome:

Warning systems use anomalies in the data to predict an event. This process requires constant data collection and comparison of trends over time. Handling this volume of data is expensive, extracting topics is elaborate, and maintaining trends for future comparison is expensive.

Malicious users and most importantly bots roar over social media and affect the discussion when organized in large groups. It is even difficult to for humans to distinguish complex bots from humans. Misleading content such as rumors is also harmful in aftermath of a crisis and finding the source of a rumor, the intention of spreading it, and intervention, before it goes viral, is laborious. 

Another area for potential improvement is ground truth acquisition for machine learning methods that automate extraction of specific posts. There have been efforts toward crowdsourcing such tasks but it is still challenging. Each disaster, location, and time has its own specificity and no global method could have been proposed which can be trained in one situation and be used in others. 

The last point is the integration of data and methods from different fields. Seismologists collect abundant amount of data from sensors and have meticulous methods for detecting earthquakes. On the other hand, the enormous amount of data is published on social media, moments after the earthquake. Integration of social media data with other sources could increase the accuracy of the information to be collected/disseminated. There are some efforts in this direction~\cite{guy2010integration,musaev2014litmus} but there is room for improvement.

%% file: sigkddExp.bbl
\begin{thebibliography}{100}

\bibitem{disasterandemergency}
Disasters and emergencies: Definitions.
\newblock \url{http://apps.who.int/disasters/repo/7656.pdf}, 2002.
\newblock accessed 09 Jan 2017.

\bibitem{url2011Twitter}
Twitter responds to the japanese disaster.
\newblock \url{https://goo.gl/8V1WC7}, Mar. 17, 2011.
\newblock accessed 13 Feb 2017.

\bibitem{url2016hurricane}
Hurricane matthew recap: Destruction from the caribbean to the united states.
\newblock \url{https://goo.gl/8Rfn6S}, Oct. 9, 2016.
\newblock accessed 10 Oct 2016.

\bibitem{allan1998topic}
J.~Allan, J.~G. Carbonell, G.~Doddington, J.~Yamron, and Y.~Yang.
\newblock Topic detection and tracking pilot study final report.
\newblock In {\em Proceedings of the DARPA Broadcast News Transcription and
  Understanding Workshop}, pages 194--218, 1998.

\bibitem{asur2010predicting}
S.~Asur and B.~A. Huberman.
\newblock Predicting the future with social media.
\newblock In {\em Web Intelligence and Intelligent Agent Technology (WI-IAT),
  2010 IEEE/WIC/ACM International Conference on}, volume~1, pages 492--499.
  IEEE, 2010.

\bibitem{atefeh2015survey}
F.~Atefeh and W.~Khreich.
\newblock A survey of techniques for event detection in twitter.
\newblock {\em Computational Intelligence}, 31(1):132--164, 2015.

\bibitem{backstrom2010find}
L.~Backstrom, E.~Sun, and C.~Marlow.
\newblock Find me if you can: improving geographical prediction with social and
  spatial proximity.
\newblock In {\em Proceedings of the 19th international conference on World
  wide web}, pages 61--70. ACM, 2010.

\bibitem{becker2011automatic}
H.~Becker, F.~Chen, D.~Iter, M.~Naaman, and L.~Gravano.
\newblock Automatic identification and presentation of twitter content for
  planned events.
\newblock In {\em ICWSM}, pages 655--656, 2011.

\bibitem{becker2011beyond}
H.~Becker, M.~Naaman, and L.~Gravano.
\newblock Beyond trending topics: Real-world event identification on twitter.
\newblock {\em ICWSM}, 11:438--441, 2011.

\bibitem{beigi2016overview}
G.~Beigi, X.~Hu, R.~Maciejewski, and H.~Liu.
\newblock An overview of sentiment analysis in social media and its
  applications in disaster relief.
\newblock In {\em Sentiment Analysis and Ontology Engineering}, pages 313--340.
  Springer, 2016.

\bibitem{blei2003latent}
D.~M. Blei, A.~Y. Ng, and M.~I. Jordan.
\newblock Latent dirichlet allocation.
\newblock {\em Journal of machine Learning research}, 3(Jan):993--1022, 2003.

\bibitem{smith2013environmental}
B.~J. Boruff.
\newblock Environmental hazards: Assessing risk and reducing disasters, 5th
  edition � by keith smith and david n. petley.
\newblock {\em Geographical Research}, 47(4):454--455, 2009.

\bibitem{bruns2012tools}
A.~Bruns and Y.~E. Liang.
\newblock Tools and methods for capturing twitter data during natural
  disasters.
\newblock {\em First Monday}, 17(4), 2012.

\bibitem{castillo2016big}
C.~Castillo.
\newblock {\em Big Crisis Data}.
\newblock Cambridge University Press, 2016.

\bibitem{chae2014public}
J.~Chae, D.~Thom, Y.~Jang, S.~Kim, T.~Ertl, and D.~S. Ebert.
\newblock Public behavior response analysis in disaster events utilizing visual
  analytics of microblog data.
\newblock {\em Computers \& Graphics}, 38:51--60, 2014.

\bibitem{cheng2010you}
Z.~Cheng, J.~Caverlee, and K.~Lee.
\newblock You are where you tweet: a content-based approach to geo-locating
  twitter users.
\newblock In {\em Proceedings of the 19th ACM international conference on
  Information and knowledge management}, pages 759--768. ACM, 2010.

\bibitem{chu2010tweeting}
Z.~Chu, S.~Gianvecchio, H.~Wang, and S.~Jajodia.
\newblock Who is tweeting on twitter: human, bot, or cyborg?
\newblock In {\em Proceedings of the 26th annual computer security applications
  conference}, pages 21--30. ACM, 2010.

\bibitem{cleveland1990stl}
R.~B. Cleveland, W.~S. Cleveland, J.~E. McRae, and I.~Terpenning.
\newblock Stl: A seasonal-trend decomposition procedure based on loess.
\newblock {\em Journal of Official Statistics}, 6(1):3--33, 1990.

\bibitem{cohn2004linguistic}
M.~A. Cohn, M.~R. Mehl, and J.~W. Pennebaker.
\newblock Linguistic markers of psychological change surrounding september 11,
  2001.
\newblock {\em Psychological science}, 15(10):687--693, 2004.

\bibitem{cordeiro2012twitter}
M.~Cordeiro.
\newblock Twitter event detection: combining wavelet analysis and topic
  inference summarization.
\newblock In {\em Doctoral Symposium on Informatics Engineering}, 2012.

\bibitem{cordeiro2016online}
M.~Cordeiro and J.~Gama.
\newblock Online social networks event detection: a survey.
\newblock In {\em Solving Large Scale Learning Tasks. Challenges and
  Algorithms}, pages 1--41. Springer, 2016.

\bibitem{crowe2012disasters}
A.~Crowe.
\newblock {\em Disasters 2.0: The application of social media systems for
  modern emergency management}.
\newblock CRC press, 2012.

\bibitem{cui2011textflow}
W.~Cui, S.~Liu, L.~Tan, C.~Shi, Y.~Song, Z.~Gao, H.~Qu, and X.~Tong.
\newblock Textflow: Towards better understanding of evolving topics in text.
\newblock {\em IEEE transactions on visualization and computer graphics},
  17(12):2412--2421, 2011.

\bibitem{difonzo2007rumor}
N.~DiFonzo and P.~Bordia.
\newblock {\em Rumor psychology: Social and organizational approaches}.
\newblock American Psychological Association, 2007.

\bibitem{dynes1970organized}
R.~R. Dynes.
\newblock {\em Organized behavior in disaster}.
\newblock Heath LexingtonBooks, 1970.

\bibitem{Ellis2015Twitter}
E.~Ellis.
\newblock How the usgs uses twitter data to track earthquakes.
\newblock \url{https://goo.gl/E6r0b2}, Oct. 7, 2015.
\newblock accessed 28 Nov 2016.

\bibitem{faulkner2014community}
M.~Faulkner, R.~Clayton, T.~Heaton, K.~M. Chandy, M.~Kohler, J.~Bunn, R.~Guy,
  A.~Liu, M.~Olson, M.~Cheng, et~al.
\newblock Community sense and response systems: Your phone as quake detector.
\newblock {\em Communications of the ACM}, 57(7):66--75, 2014.

\bibitem{faulkner2011next}
M.~Faulkner, M.~Olson, R.~Chandy, J.~Krause, K.~M. Chandy, and A.~Krause.
\newblock The next big one: Detecting earthquakes and other rare events from
  community-based sensors.
\newblock In {\em Information Processing in Sensor Networks (IPSN), 2011 10th
  International Conference on}, pages 13--24. IEEE, 2011.

\bibitem{gupta20131}
A.~Gupta, H.~Lamba, and P.~Kumaraguru.
\newblock \$1.00 per rt \#bostonmarathon \#prayforboston: Analyzing fake
  content on twitter.
\newblock In {\em eCrime Researchers Summit (eCRS)}, pages 1--12. IEEE, 2013.

\bibitem{gupta2013faking}
A.~Gupta, H.~Lamba, P.~Kumaraguru, and A.~Joshi.
\newblock Faking sandy: characterizing and identifying fake images on twitter
  during hurricane sandy.
\newblock In {\em Proceedings of the 22nd international conference on World
  Wide Web}, pages 729--736. ACM, 2013.

\bibitem{guy2010integration}
M.~Guy, P.~Earle, C.~Ostrum, K.~Gruchalla, and S.~Horvath.
\newblock Integration and dissemination of citizen reported and seismically
  derived earthquake information via social network technologies.
\newblock In {\em International Symposium on Intelligent Data Analysis}, pages
  42--53. Springer, 2010.

\bibitem{han2013stacking}
B.~Han, P.~Cook, and T.~Baldwin.
\newblock A stacking-based approach to twitter user geolocation prediction.
\newblock In {\em ACL (Conference System Demonstrations)}, pages 7--12, 2013.

\bibitem{havre2000themeriver}
S.~Havre, B.~Hetzler, and L.~Nowell.
\newblock Themeriver: Visualizing theme changes over time.
\newblock In {\em Information Visualization, 2000. InfoVis 2000. IEEE Symposium
  on}, pages 115--123. IEEE, 2000.

\bibitem{heymann2007fighting}
P.~Heymann, G.~Koutrika, and H.~Garcia-Molina.
\newblock Fighting spam on social web sites: A survey of approaches and future
  challenges.
\newblock {\em IEEE Internet Computing}, 11(6):36--45, 2007.

\bibitem{hill1962families}
R.~Hill and D.~Hansen.
\newblock Families in disaster.
\newblock In {\em Man and Society in Disaster}, pages 185--221. Basic Books,
  1962.

\bibitem{houston2015social}
J.~B. Houston, J.~Hawthorne, M.~F. Perreault, E.~H. Park, M.~Goldstein~Hode,
  M.~R. Halliwell, S.~E. Turner~McGowen, R.~Davis, S.~Vaid, J.~A. McElderry,
  et~al.
\newblock Social media and disasters: a functional framework for social media
  use in disaster planning, response, and research.
\newblock {\em Disasters}, 39(1):1--22, 2015.

\bibitem{imran2015processing}
M.~Imran, C.~Castillo, F.~Diaz, and S.~Vieweg.
\newblock Processing social media messages in mass emergency: A survey.
\newblock {\em ACM Computing Surveys (CSUR)}, 47(4):67:1--67:38, 2015.

\bibitem{imran2014aidr}
M.~Imran, C.~Castillo, J.~Lucas, P.~Meier, and S.~Vieweg.
\newblock Aidr: Artificial intelligence for disaster response.
\newblock In {\em Proceedings of the 23rd International Conference on World
  Wide Web}, pages 159--162. ACM, 2014.

\bibitem{imran2013practical}
M.~Imran, S.~Elbassuoni, C.~Castillo, F.~Diaz, and P.~Meier.
\newblock Practical extraction of disaster-relevant information from social
  media.
\newblock In {\em Proceedings of the 22nd International Conference on World
  Wide Web}, pages 1021--1024. ACM, 2013.

\bibitem{imran2013extracting}
M.~Imran, S.~M. Elbassuoni, C.~Castillo, F.~Diaz, and P.~Meier.
\newblock Extracting information nuggets from disaster-related messages in
  social media.
\newblock {\em Proc. of ISCRAM, Baden-Baden, Germany}, 2013.

\bibitem{john2009studying}
J.~P. John, A.~Moshchuk, S.~D. Gribble, A.~Krishnamurthy, et~al.
\newblock Studying spamming botnets using botlab.
\newblock In {\em NSDI}, volume~9, pages 291--306, 2009.

\bibitem{jurafsky2014speech}
D.~Jurafsky and J.~H. Martin.
\newblock {\em Speech and language processing}.
\newblock Pearson, 2014.

\bibitem{jurgens2013s}
D.~Jurgens.
\newblock That's what friends are for: Inferring location in online social
  media platforms based on social relationships.
\newblock {\em ICWSM}, 13:273--282, 2013.

\bibitem{kalyanam2015leveraging}
J.~Kalyanam, A.~Mantrach, D.~Saez-Trumper, H.~Vahabi, and G.~Lanckriet.
\newblock Leveraging social context for modeling topic evolution.
\newblock In {\em Proceedings of the 21th ACM SIGKDD International Conference
  on Knowledge Discovery and Data Mining}, pages 517--526. ACM, 2015.

\bibitem{kalyanam2016event}
S.~V. M.~C. Kalyanam, Janani and G.~Lanckriet.
\newblock From event detection to story telling on microblogs.
\newblock In {\em Proceedings of the ACM/IEEE Conference on Advances in Social
  Network Analysis and Mining (ASONAM)}, pages 437--442. ACM, 2016.

\bibitem{kanver2012hurricane}
L.~Kavner.
\newblock Hurricane sandy: Red cross, other relief organizations see social
  media as ‘double-edged sword’ for relief efforts.
\newblock \url{https://goo.gl/angXF8}, Oct. 31, 2012.
\newblock accessed 09 Jan 2017.

\bibitem{koh2014only}
Y.~Koh.
\newblock Only 11\% of new twitter users in 2012 are still tweeting.
\newblock \url{https://goo.gl/vl9D3h}, Mar. 21, 2014.
\newblock accessed 13 Feb 2017.

\bibitem{kossinets2006empirical}
G.~Kossinets and D.~J. Watts.
\newblock Empirical analysis of an evolving social network.
\newblock {\em science}, 311(5757):88--90, 2006.

\bibitem{kullback1951information}
S.~Kullback and R.~A. Leibler.
\newblock On information and sufficiency.
\newblock {\em The annals of mathematical statistics}, 22(1):79--86, 1951.

\bibitem{kumar2011tweettracker}
S.~Kumar, G.~Barbier, M.~A. Abbasi, and H.~Liu.
\newblock Tweettracker: An analysis tool for humanitarian and disaster relief.
\newblock In {\em ICWSM}, pages 661--662, 2011.

\bibitem{kumar2014behavior}
S.~Kumar, X.~Hu, and H.~Liu.
\newblock A behavior analytics approach to identifying tweets from crisis
  regions.
\newblock In {\em Proceedings of the 25th ACM conference on Hypertext and
  social media}, pages 255--260. ACM, 2014.

\bibitem{lee2001algorithms}
D.~D. Lee and H.~S. Seung.
\newblock Algorithms for non-negative matrix factorization.
\newblock In {\em Advances in neural information processing systems}, pages
  556--562, 2001.

\bibitem{lee2011seven}
K.~Lee, B.~D. Eoff, and J.~Caverlee.
\newblock Seven months with the devils: A long-term study of content polluters
  on twitter.
\newblock In {\em ICWSM}, pages 185--192, 2011.

\bibitem{lehmann2012dynamical}
J.~Lehmann, B.~Gon{\c{c}}alves, J.~J. Ramasco, and C.~Cattuto.
\newblock Dynamical classes of collective attention in twitter.
\newblock In {\em Proceedings of the 21st international conference on World
  Wide Web}, pages 251--260. ACM, 2012.

\bibitem{leskovec2009meme}
J.~Leskovec, L.~Backstrom, and J.~Kleinberg.
\newblock Meme-tracking and the dynamics of the news cycle.
\newblock In {\em Proceedings of the 15th ACM SIGKDD international conference
  on Knowledge discovery and data mining}, pages 497--506. ACM, 2009.

\bibitem{li2012towards}
R.~Li, S.~Wang, H.~Deng, R.~Wang, and K.~C.-C. Chang.
\newblock Towards social user profiling: unified and discriminative influence
  model for inferring home locations.
\newblock In {\em Proceedings of the 18th ACM SIGKDD international conference
  on Knowledge discovery and data mining}, pages 1023--1031. ACM, 2012.

\bibitem{luckerson2014fear}
V.~Luckerson.
\newblock Fear, misinformation, and social media complicate ebola fight.
\newblock {\em Time Magazine}, 2014.

\bibitem{mahmud2012tweet}
J.~Mahmud, J.~Nichols, and C.~Drews.
\newblock Where is this tweet from? inferring home locations of twitter users.
\newblock {\em ICWSM}, 12:511--514, 2012.

\bibitem{mathioudakis2010twittermonitor}
M.~Mathioudakis and N.~Koudas.
\newblock Twittermonitor: trend detection over the twitter stream.
\newblock In {\em Proceedings of the 2010 ACM SIGMOD International Conference
  on Management of data}, pages 1155--1158. ACM, 2010.

\bibitem{meier2015digital}
P.~Meier.
\newblock {\em Digital humanitarians: how big data is changing the face of
  humanitarian response}.
\newblock Crc Press, 2015.

\bibitem{meier2012crisis}
P.~Meier.
\newblock How crisis mapping saved lives in haiti.
\newblock \url{https://goo.gl/uASbu8}, Jul. 2, 2012.
\newblock accessed 09 Jan 2017.

\bibitem{mendoza2010twitter}
M.~Mendoza, B.~Poblete, and C.~Castillo.
\newblock Twitter under crisis: can we trust what we rt?
\newblock In {\em Proceedings of the first workshop on social media analytics},
  pages 71--79. ACM, 2010.

\bibitem{mishne2006predicting}
G.~Mishne, N.~S. Glance, et~al.
\newblock Predicting movie sales from blogger sentiment.
\newblock In {\em AAAI spring symposium: computational approaches to analyzing
  weblogs}, pages 155--158, 2006.

\bibitem{morstatter2014finding}
F.~Morstatter, N.~Lubold, H.~Pon-Barry, J.~Pfeffer, and H.~Liu.
\newblock Finding eyewitness tweets during crises.
\newblock {\em arXiv preprint arXiv:1403.1773}, 2014.

\bibitem{morstatter2016new}
F.~Morstatter, L.~Wu, T.~H. Nazer, K.~M. Carley, and H.~Liu.
\newblock A new approach to bot detection: striking the balance between
  precision and recall.
\newblock In {\em Proceedings of the IEEE/ACM International Conference on
  Advances in Social Networks Analysis and Mining (ASONAM)}, pages 533--540,
  2016.

\bibitem{musaev2014litmus}
A.~Musaev, D.~Wang, and C.~Pu.
\newblock Litmus: Landslide detection by integrating multiple sources.
\newblock In {\em 11th International Conference Information Systems for Crisis
  Response and Management (ISCRAM)}, pages 677--686, 2014.

\bibitem{naaman2011hip}
M.~Naaman, H.~Becker, and L.~Gravano.
\newblock Hip and trendy: Characterizing emerging trends on twitter.
\newblock {\em Journal of the American Society for Information Science and
  Technology}, 62(5):902--918, 2011.

\bibitem{niculae2015quotus}
V.~Niculae, C.~Suen, J.~Zhang, C.~Danescu-Niculescu-Mizil, and J.~Leskovec.
\newblock Quotus: The structure of political media coverage as revealed by
  quoting patterns.
\newblock In {\em Proceedings of the 24th International Conference on World
  Wide Web}, pages 798--808. ACM, 2015.

\bibitem{okolloh2009ushahidi}
O.~Okolloh.
\newblock Ushahidi, or ‘testimony’: Web 2.0 tools for crowdsourcing crisis
  information.
\newblock {\em Participatory learning and action}, 59(1):65--70, 2009.

\bibitem{olteanu2014crisislex}
A.~Olteanu, C.~Castillo, F.~Diaz, and S.~Vieweg.
\newblock Crisislex: A lexicon for collecting and filtering microblogged
  communications in crises.
\newblock In {\em ICWSM}, pages 376--385, 2014.

\bibitem{palen2016crisis}
L.~Palen and K.~M. Anderson.
\newblock Crisis informatics—new data for extraordinary times.
\newblock {\em Science}, 353(6296):224--225, 2016.

\bibitem{Palen224}
L.~Palen and K.~M. Anderson.
\newblock Crisis informatics{\textemdash}new data for extraordinary times.
\newblock {\em Science}, 353(6296):224--225, 2016.

\bibitem{palm2016hurricane}
B.~Palm.
\newblock Hurricane matthew reaches category 4 status, barreling toward
  florida.
\newblock \url{https://goo.gl/ZW33U3}, Oct. 6, 2016.
\newblock accessed 10 Oct 2016.

\bibitem{popescu2010detecting}
A.-M. Popescu and M.~Pennacchiotti.
\newblock Detecting controversial events from twitter.
\newblock In {\em Proceedings of the 19th ACM international conference on
  Information and knowledge management}, pages 1873--1876. ACM, 2010.

\bibitem{popescu2011extracting}
A.-M. Popescu, M.~Pennacchiotti, and D.~Paranjpe.
\newblock Extracting events and event descriptions from twitter.
\newblock In {\em Proceedings of the 20th international conference companion on
  World wide web}, pages 105--106. ACM, 2011.

\bibitem{powell1954introduction}
J.~W. Powell.
\newblock An introduction to the natural history of disaster.
\newblock {\em Univ. of Maryland: Disaster Research Project}, 1954.

\bibitem{power2014emergency}
R.~Power, B.~Robinson, J.~Colton, and M.~Cameron.
\newblock Emergency situation awareness: Twitter case studies.
\newblock In {\em International Conference on Information Systems for Crisis
  Response and Management in Mediterranean Countries}, pages 218--231.
  Springer, 2014.

\bibitem{purohit2013emergency}
H.~Purohit, C.~Castillo, F.~Diaz, A.~Sheth, and P.~Meier.
\newblock Emergency-relief coordination on social media: Automatically matching
  resource requests and offers.
\newblock {\em First Monday}, 19(1), 2013.

\bibitem{ratkiewicz2011truthy}
J.~Ratkiewicz, M.~Conover, M.~Meiss, B.~Gon{\c{c}}alves, S.~Patil, A.~Flammini,
  and F.~Menczer.
\newblock Truthy: mapping the spread of astroturf in microblog streams.
\newblock In {\em Proceedings of the 20th international conference companion on
  World wide web}, pages 249--252. ACM, 2011.

\bibitem{reese2016how}
A.~Reese.
\newblock How we'll predict the next natural disaster: Advances in natural
  hazard forecasting could help keep more people out of harm's way.
\newblock {\em Discover Magazine}, Sep. 2016.

\bibitem{sakaki2010earthquake}
T.~Sakaki, M.~Okazaki, and Y.~Matsuo.
\newblock Earthquake shakes twitter users: real-time event detection by social
  sensors.
\newblock In {\em Proceedings of the 19th international conference on World
  wide web}, pages 851--860. ACM, 2010.

\bibitem{Sambuli2013useful}
N.~SAMBULI.
\newblock How useful is a tweet? a review of the first tweets from the westgate
  mall attack.
\newblock \url{https://goo.gl/qRGYZD}, Oct. 3, 2013.
\newblock accessed 10 Feb 2017.

\bibitem{sampson2015real}
J.~Sampson, F.~Morstatter, R.~Zafarani, and H.~Liu.
\newblock Real-time crisis mapping using language distribution.
\newblock In {\em 2015 IEEE International Conference on Data Mining Workshop
  (ICDMW)}, pages 1648--1651. IEEE, 2015.

\bibitem{aim1986schramm}
D.~Schramm and R.~Hansen.
\newblock Aim \& scope of disaster management: Study guide and course text.
\newblock {\em Disaster Management Center, University of Wisconsin-Madison},
  1986.

\bibitem{schulz2013multi}
A.~Schulz, A.~Hadjakos, H.~Paulheim, J.~Nachtwey, and M.~M{\"u}hlh{\"a}user.
\newblock A multi-indicator approach for geolocalization of tweets.
\newblock In {\em ICWSM}, pages 573--582, 2013.

\bibitem{sederholm2015stranded}
J.~SEDERHOLM.
\newblock \#strandedinus: Americans open homes to strangers stuck after paris
  attacks.
\newblock \url{https://goo.gl/NQqEDh}, Nov. 14, 2015.
\newblock accessed 09 Jan 2017.

\bibitem{sorensen2000hazard}
J.~H. Sorensen.
\newblock Hazard warning systems: Review of 20 years of progress.
\newblock {\em Natural Hazards Review}, 1(2):119--125, 2000.

\bibitem{starbird2010tweak}
K.~Starbird and J.~Stamberger.
\newblock Tweak the tweet: Leveraging microblogging proliferation with a
  prescriptive syntax to support citizen reporting.
\newblock In {\em Proceedings of the 7th International ISCRAM
  Conference--Seattle}, pages 1--5, 2010.

\bibitem{szabo2010predicting}
G.~Szabo and B.~A. Huberman.
\newblock Predicting the popularity of online content.
\newblock {\em Communications of the ACM}, 53(8):80--88, 2010.

\bibitem{temnikova2015emterms}
I.~Temnikova, C.~Castillo, and S.~Vieweg.
\newblock Emterms 1. 0: a terminological resource for crisis tweets.
\newblock In {\em ISCRAM 2015 proceedings of the 12th international conference
  on information systems for crisis response and management}, 2015.

\bibitem{thomas2012adapting}
K.~Thomas, C.~Grier, and V.~Paxson.
\newblock Adapting social spam infrastructure for political censorship.
\newblock In {\em LEET}, 2012.

\bibitem{tumasjan2010predicting}
A.~Tumasjan, T.~O. Sprenger, P.~G. Sandner, and I.~M. Welpe.
\newblock Predicting elections with twitter: What 140 characters reveal about
  political sentiment.
\newblock {\em ICWSM}, 10:178--185, 2010.

\bibitem{Vaca2014time}
C.~K. Vaca, A.~Mantrach, A.~Jaimes, and M.~Saerens.
\newblock A time-based collective factorization for topic discovery and
  monitoring in news.
\newblock In {\em Proceedings of the 23rd International Conference on World
  Wide Web}, WWW '14, pages 527--538, New York, NY, USA, 2014. ACM.

\bibitem{varga2013aid}
I.~Varga, M.~Sano, K.~Torisawa, C.~Hashimoto, K.~Ohtake, T.~Kawai, J.-H. Oh,
  and S.~De~Saeger.
\newblock Aid is out there: Looking for help from tweets during a large scale
  disaster.
\newblock In {\em ACL (1)}, pages 1619--1629, 2013.

\bibitem{vieweg2010microblogging}
S.~Vieweg, A.~L. Hughes, K.~Starbird, and L.~Palen.
\newblock Microblogging during two natural hazards events: what twitter may
  contribute to situational awareness.
\newblock In {\em Proceedings of the SIGCHI conference on human factors in
  computing systems}, pages 1079--1088. ACM, 2010.

\bibitem{wang2012automatic}
X.~Wang, M.~S. Gerber, and D.~E. Brown.
\newblock Automatic crime prediction using events extracted from twitter posts.
\newblock In {\em International Conference on Social Computing,
  Behavioral-Cultural Modeling, and Prediction}, pages 231--238. Springer,
  2012.

\bibitem{wolpert1992stacked}
D.~H. Wolpert.
\newblock Stacked generalization.
\newblock {\em Neural networks}, 5(2):241--259, 1992.

\bibitem{wu2015social}
F.~Wu, J.~Shu, Y.~Huang, and Z.~Yuan.
\newblock Social spammer and spam message co-detection in microblogging with
  social context regularization.
\newblock In {\em Proceedings of the 24th ACM International on Conference on
  Information and Knowledge Management}, pages 1601--1610. ACM, 2015.

\bibitem{xie2013topicsketch}
W.~Xie, F.~Zhu, J.~Jiang, E.-P. Lim, and K.~Wang.
\newblock Topicsketch: Real-time bursty topic detection from twitter.
\newblock In {\em 2013 IEEE 13th International Conference on Data Mining},
  pages 837--846. IEEE, 2013.

\bibitem{xie2008spamming}
Y.~Xie, F.~Yu, K.~Achan, R.~Panigrahy, G.~Hulten, and I.~Osipkov.
\newblock Spamming botnets: signatures and characteristics.
\newblock {\em ACM SIGCOMM Computer Communication Review}, 38(4):171--182,
  2008.

\bibitem{yu2012survey}
S.~Yu and S.~Kak.
\newblock A survey of prediction using social media.
\newblock {\em arXiv preprint arXiv:1203.1647}, 2012.

\bibitem{zafarani201510}
R.~Zafarani and H.~Liu.
\newblock 10 bits of surprise: Detecting malicious users with minimum
  information.
\newblock In {\em Proceedings of the 24th ACM International on Conference on
  Information and Knowledge Management}, pages 423--431. ACM, 2015.

\bibitem{zhang2009improving}
W.~Zhang and S.~Skiena.
\newblock Improving movie gross prediction through news analysis.
\newblock In {\em Proceedings of the 2009 IEEE/WIC/ACM International Joint
  Conference on Web Intelligence and Intelligent Agent Technology-Volume 01},
  pages 301--304. IEEE Computer Society, 2009.

\bibitem{zook2010volunteered}
M.~Zook, M.~Graham, T.~Shelton, and S.~Gorman.
\newblock Volunteered geographic information and crowdsourcing disaster relief:
  a case study of the haitian earthquake.
\newblock {\em World Medical and Health Policy by Wiley Online Library},
  2(2):7--33, 2010.

\end{thebibliography}
